# A Study of the Grunwald-Letnikov Definition for Minimizing the Effects of Random Noise on Fractional Order Differential Equations

Mithun Chakraborty, Deepyaman Maiti, Amit Konar
Dept. of Electronics and Telecommunication Engineering,
Jadavpur University,
Kolkata, India.
mithun.chakra108@gmail.com,
deepyamanmaiti@gmail.com, konaramit@yahoo.co.in

Ramadoss Janarthanan
Dept. of Information Technology,
Jaya Engineering College,
Chennai, India.
srmjana_73@yahoo.com

*Abstract*—Of the many definitions for fractional order differintegral, the Grunwald-Letnikov definition is arguably the most important one. The necessity of this definition for the description and analysis of fractional order systems cannot be overstated. Unfortunately, the Fractional Order Differential Equation (FODE) describing such a systems, in its original form, highly sensitive to the effects of random noise components inevitable in a natural environment. Thus direct application of the definition in a real-life problem can yield erroneous results. In this article, we perform an in-depth mathematical analysis the Grunwald-Letnikov definition in depth and, as far as we know, we are the first to do so. Based on our analysis, we present a transformation scheme which will allow us to accurately analyze generalized fractional order systems in presence of significant quantities of random errors. Finally, by a simple experiment, we demonstrate the high degree of robustness to noise offered by the said transformation and thus validate our scheme.

## I. INTRODUCTION

The first references to fractional order derivatives were made in the 17$^{th}$ century. Since then, the theory of fractional calculus has been highly developed by many mathematicians. In the last five decades, many authors made a great effort to apply this knowledge in practice. The real-world objects or processes that we need to deal with are generally of fractional order [1]. A typical example of a non-integer (fractional) order system is the voltage-current relation of a semi-infinite lossy RC line or diffusion of heat into a semi-infinite solid, where heat flow q(t) is equal to the half-derivative of temperature T(t) [2]:

$$\frac{d^{0.5}T(t)}{dt^{0.5}} = q(t). \qquad (1)$$

The usual practice when dealing with such a fractional order process had been to use an integer order approximation. The fractional order of the system was ignored because of the non-existence of simple mathematical tools for the description of such systems. Since major advances have been made in this area recently, it is possible to consider also the real order of the dynamical systems with distributed parameters.

The first generalizations of fractional order systems such as s-plane representation, root locus and frequency response were made in pioneering works such as [3]. Methods for the state-space analysis of fractional order processes were proposed in [4], [5], where possible schemes of deriving a state-space model from a transfer function are discussed in depth. In this context, polynomial expression and continued fraction expression (CFE) of Tustin's operators are two very useful tools.

Researches in fractional order controllers and fractional order system identification have produced a large number of achievements. Podlubny [6] proposed and demonstrated the concept of the fractional order PI$^\lambda$D$^\delta$ controllers. The representative design methods of such controllers are found in [7] – [9]. Computation of transfer characteristics of the fractional order dynamic systems has been the subject of several publications: by numerical as well as by analytical methods. Related schemes and applications can be viewed in [10] – [14].

It must be appreciated that all the above applications in the field of fractional order control systems must utilize the basic definitions of fractional differintegral. The two most useful of such definitions are those by Riemann-Liouville and Grunwald-Letnikov [15].

The published articles on fractional order systems, that we could find, assumed or dealt with an ideal scenario. For example, in a scheme for system identification that used the output readings from a control system, all the (synthetic) readings used were 100% accurate, whereas a real-life scenario could have been implemented by adding a random error component to each such synthetic reading.

This observation set us to investigate the effects of random error components or natural noise superimposed over the ideal data. If effects resulting from random noise are included in the analysis of a fractional order system, the results of the analysis will be different each time because of the randomness. If the said effect is large enough, the results will be so unpredictable as to be useless. Our research focuses on studying the most important of the differintegral definitions, the one by Grunwald-Letnikov, with an intent to find an algorithm to minimize the effects of natural noise. We will look at some control theory applications too, for our ultimate aim is to accurately describe a fractional order system in presence of random errors and fluctuations.

## II. THE GRUNWALD-LETNIKOV DEFINITION

We will recapitulate, briefly, some of the important results concerning the Grunwald-Letnikov definition of the fractional differintegral. At first, we generalize the differential and integral operators into one fundamental operator $_aD_t^\alpha$ where:

$$_aD_t^\alpha = \begin{cases} \dfrac{d^\alpha}{dt^\alpha}, & \Re(\alpha) > 0 \\ 1, & \Re(\alpha) = 0 \\ \int_a^t (d\tau)^{-\alpha}, & \Re(\alpha) < 0 \end{cases} \qquad (2)$$

$\Re(\alpha)$ denotes the real part of $\alpha$ which is a complex quantity, in general.

The Grunwald-Letnikov definition of the fractional differintegral of order $\alpha$ of a time-function $f(t)$ at time-instant $t$ is:

$$_aD_t^\alpha f(t) = \lim_{h \to 0} \frac{1}{h^\alpha} \sum_{j=0}^{\left[\frac{t-a}{h}\right]} b_j f(t - jh) \qquad (3)$$

where $[y]$ stands for the greatest integer not exceeding $y$ and the weighting coefficients $b_j$ can be calculated recursively by the following formulae:

$$b_0 = 1, \quad b_j = \left(1 - \frac{1+\alpha}{j}\right) b_{j-1}, \quad (j \geq 1) \qquad (4)$$

$$\Rightarrow b_1 = (-\alpha) b_0 = -\alpha, \quad b_2 = \frac{(1-\alpha)}{2} b_1 = \frac{(-\alpha)(1-\alpha)}{2}, \ldots \qquad (5)$$

In general,

$$b_n = \frac{1}{n!}(-\alpha)(1-\alpha)(2-\alpha)\ldots\{(n-1)-\alpha\}, \quad n=0,1,2,\ldots \qquad (6)$$

Derived from the Grunwald-Letnikov definition, the formula for numerical computation of the fractional differintegral is:

$$_{t-L}D_t^\alpha f(t) \approx h^{-\alpha} \sum_{j=0}^{\lfloor L/T \rfloor} b_j f(t-jh) = \sum_{j=0}^N \Phi_j(\alpha) f(t-jh) \qquad (7)$$

where L is the length of memory and N = [L/T]; $T$ ($\ll 1$ s), the sampling time always replaces the time increment h during approximation. The right hand side of (7) is nothing but the weighted sum of a very large number of terms, the weight of the $(t - jh)^{th}$ term being $\Phi_j(\alpha) = h^{-\alpha} b_j$.

Let us call the function
$$\Phi_n(\alpha) = \frac{h^{-\alpha}}{n!}(-\alpha)(1-\alpha)(2-\alpha)\ldots\{(n-1)-\alpha\}, \quad n=0,1,2,\ldots \quad 0<h<1,$$
$-\infty < \alpha < \infty$ the Grunwald-Letnikov function of degree n. The expression for $\Phi_n(\alpha)$ has been derived by us for the purpose of this work, as will be elucidated subsequently, and is not seen in previous literature. We are also introducing the term "Grunwald-Letnikov function of degree n" for this article.

## III. DESCRIPTION OF FRACTIONAL ORDER SYSTEMS

Fractional order systems are described in time domain by a system of fractional order differential equations (FODE) [21] such as:

$$a_n D^{\alpha_n} c(t) + a_{n-1} D^{\alpha_{n-1}} c(t) + a_{n-2} D^{\alpha_{n-2}} c(t) \\ + \ldots\ldots + a_0 c(t) = r(t) \qquad (8)$$

where r(t) is the system input, c(t) is the system output.

The important point to be noted here is that the orders of differentiation $\alpha_n, \alpha_{n-1}, \ldots \alpha_1$ are fractional. That is, $\alpha_n, \alpha_{n-1}, \ldots \alpha_1$ can assume any real values and are not restricted to only integers. In contrast to the integer order differential equations which are well studied, analytical solution of FODEs is rather complicated, and for higher order equations, almost impossible. To circumvent this difficulty, various methods have been devised to obtain numerical solutions. These numerical solutions usually make use of the Grunwald-Letnikov definition for fractional order differintegral.

In the real world, presence of random noise components affects the accuracy of the differential equations that describe a system. Consider for example the following fractional order system with a transfer function: $\dfrac{1}{a_1 s^{2.23} + a_2 s^{0.88} + a_3}$, $a_1$, $a_2$ and $a_3$ being arbitrary constants.

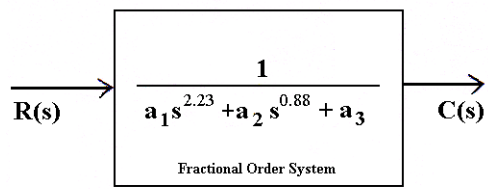

Figure 1. A generic fractional order system.

The FODE describing this system is:

$$r(t) = a_1 D^{2.23} c(t) + a_2 D^{0.88} c(t) + a_3 c(t). \qquad (9)$$

The problem of system identification then reduces to determination of the system parameters $a_1$, $a_2$, $a_3$ based on our measurement of the system output for a given input (such as unit step signal). Our key idea is to first choose a particular t, say $t_0$. Then we substitute the values of r(t), $D^{2.23}c(t)$, $D^{0.08}c(t)$, and c(t), measured/numerically evaluated at $t = t_0$, in (9) to obtain an equation in three unknown quantities. Next we integrate both sides of (9) once and substitute the appropriate signal values computed at $t = t_0$; in the third step, we integrate both sides of (9) twice and make the necessary substitutions. We thus end up with three independent equations and their simultaneous solution yields estimates of $a_1$, $a_2$, $a_3$.

Now suppose, the output waveform gets distorted owing to random environmental fluctuations, human error etc. Considering the system to be *pure*, this distortion can be accounted for by a random error waveform e(t) superimposed on the *actual* output waveform c(t). This is equivalent to subjecting (9) to the transformation $c(t) \rightarrow c(t) + e(t)$. (10)

Then (9) should be replaced by

$$r(t) = a_1 D^{2.23}[c(t)+e(t)] + a_2 D^{0.88}[c(t)+e(t)] + a_3[c(t)+e(t)] \quad (11)$$

Thus, when we sample the *output* waveform for numerically evaluating the fractional differintegrals involved, we actually get samples of $[c(t) + e(t)]$. So (11) will not give an accurate relation between $a_1$, $a_2$ and $a_3$ due to the presence of the terms $a_1 D^{2.23}e(t)$, $a_2 D^{0.88}e(t)$ and $a_3 e(t)$. Our aim will be to minimize this inaccuracy by reducing the effects of the error waveform as far as possible. In other words, we will present a description that better represents the system in presence of error.

IV. ILLUSTRATION OF INADEQUACY OF PRESENT METHODS

To better visualize how the presence of e(t) gives a false description of a system, let us do an identification of the system in Fig. 1, in absence and in presence of noise, using the method proposed in [10] and further developed in [16]. We will identify the parameters $a_1$, $a_2$ and $a_3$. The input considered is r(t) = 1 i.e. unit step. Synthetic data for c(t) is created using $a_1 = 0.8$, $a_2 = 0.5$ and $a_3 = 1$. Length of memory L = 10 seconds and T = 0.001 seconds is used to calculate the fractional derivatives.

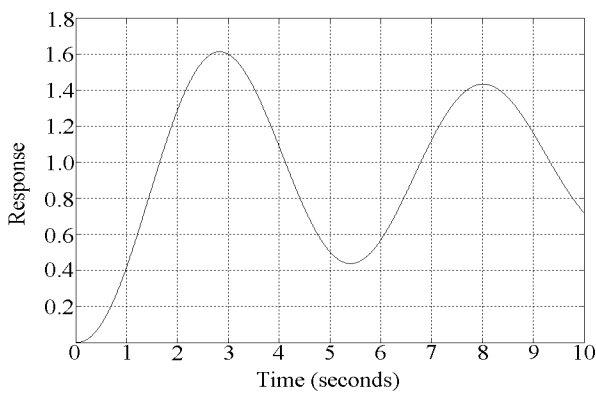

Figure 2. Unit step response c(t) used for identification.

The output waveform c(t) used to make the identification is given in Fig. 2. When no error component is added, the unknown parameters are calculated as $a_1 = 0.8033$, $a_2 = 0.4922$, $a_3 = 1.0001$. The errors in estimating them are respectively 0.4125%, 0.1600% and 0.0100%. So the identification is reasonably accurate.

Then we added a random waveform e(t) to c(t) and repeated the method, each element in the error waveform being in the range $[-0.05, 0.05]$. Now the unknown parameters are calculated as $a_1 = 0.0000$, $a_2 = 0.6248$, $a_3 = 0.9687$. This identification is obviously way off the mark, and we conclude that the method performs well under ideal conditions, but falters in a realistic scenario.

This indicates that the FODEs describing the system are quite inaccurate in presence of the random noise component. In what follows, we will propose alternative FODEs to describe the system, equations which will yield an accurate description even in presence of noise.

V. A PHILOSOPHICAL APPROACH TO THE SOLUTION

To strike the problem at its root, it is intuitively understood that the fractional differential equation viz. (8), describing the system dynamics, must be subjected to a suitable transformation that somehow eliminates (or at least minimizes) the effect of random error e(t) but magnifies (or at least preserves) the effect of actual system output c(t) on the estimation of system parameters. As the time-functions r(t) and c(t) are known by measurements or otherwise, numerical differentiation/numerical integration of both sides of (8) with respect to time are two linear transformations that immediately come to mind.

However, we know that the physical significance of time-derivative of a function at a point is that it gives the slope of the function vs. time curve at that point whereas definite integral of a function from one time-point to another is the algebraic area under the curve between those two points. Now, for the type of system we are interested in, the actual output waveform is continuous, more or less smooth, and always non-negative so that its derivative can never be expected to be very large whereas its integral is likely to be appreciable. On the other hand, the error waveform has rapid fluctuations (sharp bends) and assumes both positive and negative values so that its derivatives at certain points may work out to a numerically large value whereas an integration operation over this error waveform can be expected to yield quite a low value, since the areas with opposing signs should nullify each other in the long run. Thus, integration appears to be a viable option for serving our purpose.

Unfortunately, all this discussion is valid for integer-order differentiation/integration whereas, on integrating both sides of (8), we shall end up with fractional differintegrals in general. For the fractional order case, the Grunwald-Letnikov definition for numerical computation is essentially the same for both differentiation and integration, the only significant difference being in the sign of the order of differintegration. Thus, it is not obvious whether a similar physical significance could be attached to fractional differintegral and whether the same conclusion could be drawn about it.

In other words, we must ascertain the dependence of $D^\alpha x(t)$ on $\alpha$ for $x(t) = c(t)$ or $e(t)$ for a given t and this calls for a detailed study of the Grunwald-Letnikov function

$$\Phi_n(\alpha) = \frac{h^{-\alpha}}{n!}(-\alpha)(1-\alpha)(2-\alpha)\ldots\{(n-1)-\alpha\}, \; n=0,1,2,\ldots,$$
$$0 < h < 1, \; -\infty < \alpha < \infty.$$

## VI. ANALYSIS OF GRUNWALD-LETNIKOV FUNCTIONS $\Phi_N(\alpha)$ FOR $\alpha < 0$ AND $\alpha > 0$

For $n = 0$, $\Phi_0(\alpha) = h^{-\alpha}$ which, for $0 < h < 1$, is nothing but a rising exponential function and has a well-known nature:

$\Phi_0(\alpha)$ is strictly monotonically increasing from $-\infty$ at $\alpha \to -\infty$ to $+\infty$ at $\alpha \to +\infty$.

$\Phi_0(0) = 1$.

We are actually interested in the cases where $n \geq 1$. Some of the concepts used in the following analysis are quite standard and can be found in textbooks like [17].

### A. $\alpha < 0$: Fractional Order Integration

We put $(-\alpha) = \mu$. Let $f_n(\mu) \equiv \Phi_n(\alpha)$ for any given $n \geq 1$. Then,

$$f_n(\mu) = \frac{h^\mu}{n!}\mu(1+\mu)(2+\mu)\ldots\{(n-1)+\mu\} = \frac{h^\mu}{n!}g_n(\mu) \quad (12)$$

where $g_n(\mu) = \mu(1+\mu)(2+\mu)\ldots\{(n-1)+\mu\}$.

Differentiating $g_n(\mu)$ with respect to $\mu$,

$g'_n(\mu) = (1+\mu)(2+\mu)(3+\mu)\ldots\{(n-1)+\mu\} + \mu(2+\mu)(3+\mu)\ldots\{(n-1)+\mu\}$
$\quad + \mu(1+\mu)(3+\mu)\ldots\{(n-1)+\mu\} + \ldots\ldots$ to n terms
$= g_n(\mu)\Psi_n(\mu)$ \hfill (13)

for $\Psi_n(\mu) = \frac{1}{\mu} + \frac{1}{1+\mu} + \frac{1}{2+\mu} + \ldots + \frac{1}{(n-1)+\mu}$.

Thus,

$f'_n(\mu) = \frac{h^\mu \ln(h)}{n!}g_n(\mu) + \frac{h^\mu}{n!}g'_n(\mu)$, from (12).

$= \frac{h^\mu}{n!}[g_n(\mu)\ln(h) + g_n(\mu)\Psi_n(\mu)]$, from (13).

$= \frac{h^\mu}{n!}g_n(\mu)[-\gamma + \Psi_n(\mu)]$,

where, $\gamma = -\ln(h) > 0$ as $0 < h < 1$.

$\Rightarrow f'_n(\mu) = f_n(\mu)[-\gamma + \Psi_n(\mu)]$ \hfill (14)

Obviously, $\Psi_n(\mu) \to \infty$ as $\mu \to 0$; $\Psi_n(\mu) \to 0$ as $\mu \to \infty$; $\Psi_n(\mu)$ is a strictly monotonically decreasing function of $\mu$ for $0 < \mu < \infty$.

Thus, over the interval $0 < \mu < \infty$, $\Psi_n(\mu)$ assumes every possible positive real value once and only once.

So, for any given n, there exists one and only one positive value of $\mu$ (dependent on n), say $\mu_{m,n}$, such that $\Psi_n(\mu_{m,n}) = \gamma$ (as $\gamma > 0$), i.e. $f'_n(\mu_{m,n}) = 0$.

Now, evidently, $f_n(0) = 0$ for $n \geq 1$ and $f_n(\mu) > 0$ $\forall \; \mu > 0$. \hfill (15)

Again, as $\Psi_n(\mu)$ is monotonically decreasing for positive $\mu$ and $\Psi_n(\mu) = \gamma$ at $\mu = \mu_{m,n}$, we must have

$\Psi_n(\mu) > \gamma$ for $0 < \mu < \mu_{m,n}$
and $\Psi_n(\mu) < \gamma$ for $\mu > \mu_{m,n}$. \hfill (16)

Thus, (14), (15) and (16) imply that $f'_n(\mu) > 0$ for $0 < \mu < \mu_{m,n}$ and $f'_n(\mu) < 0$ for $\mu > \mu_{m,n}$. \hfill (17)

Moreover,

$$\lim_{\mu \to \infty} f_n(\mu) = \frac{1}{n!}\lim_{\mu \to \infty}\frac{g_n(\mu)}{h^{-\mu}},$$

where $g_n(\mu) \to \infty$ as $\mu \to \infty$ and,

also, $h^{-\mu} \to \infty$ as $\mu \to \infty$ since $0 < h < 1$.

Applying L'Hospital's Rule successively n times, it can be shown that $\lim_{\mu \to \infty} f_n(\mu) = 0.$ \hfill (18)

From (15), (17) and (18), we conclude that, for $0 \leq \mu < \infty$, $f_n(\mu)$ rises strictly monotonically from zero at $\mu = 0$ to a local maximum at $\mu = \mu_{m,n}$, then decreases strictly monotonically with further increase in $\mu$ beyond $\mu_{m,n}$ and asymptotically approaches zero. Thus, for non-negative $\mu$, the lower bound of $f_n(\mu)$ is 0 and its upper bound is $f_n(\mu_{m,n})$.

To estimate $\mu_{m,n}$, we first try to solve, for various possible values of n, the equation: $\Psi_n(\mu) = \gamma$. \hfill (19)

For $n = 1$, the equation (19) reduces to $\frac{1}{\mu} = \gamma$ which has only one root at: $\mu = 1/\gamma$. \hfill (20)

For $n = 2$, the equation (19) reduces to:

$\gamma\mu^2 + (\gamma - 2)\mu - 1 = 0$ with exactly one positive root at

$$\mu = \frac{1}{\gamma} + \left(\sqrt{\frac{1}{4} + \frac{1}{\gamma^2}} - \frac{1}{2}\right) > \frac{1}{\gamma}. \quad (21)$$

In general, for $n > 2$, we observe that

$$\Psi_n\left(\frac{1}{\gamma}\right) = \gamma + \left[\frac{1}{1+1/\gamma} + \frac{1}{2+1/\gamma} + \ldots + \frac{1}{(n-1)+1/\gamma}\right]$$

$$> \gamma \quad \text{as } \gamma > 0 \tag{22}$$

For $\mu < 1/\gamma$,

$$\Psi_n(\mu) = \frac{1}{1+\mu} + \frac{1}{2+\mu} + \ldots\ldots + \frac{1}{(n-1)+\mu}$$

$$> \gamma + \left[ \frac{1}{1+1/\gamma} + \frac{1}{2+1/\gamma} + \ldots\ldots + \frac{1}{(n-1)+1/\gamma} \right]$$

$$> \gamma \quad \text{as } \gamma > 0 \tag{23}$$

But, as already shown, for any given $n \geq 1$, there exists exactly one value of $\mu$, viz. $\mu_{m,n}$, such that:

$$\Psi_n(\mu_{m,n}) = \gamma. \tag{24}$$

Thus, (20) indicates that $\mu_{m,1} = 1/\gamma$ and (21) – (24) show that $\mu_{m,n} > 1/\gamma$ for $n > 1$.

Thus, $h^{\mu_{m,n}} \leq h^{1/\gamma}, n \geq 1$ as $0 < h < 1$.

$$\Rightarrow h^{\mu_{m,n}} \leq e^{-1}$$

as $\gamma = -\ln(h) \Rightarrow h = e^{-1/\gamma} \Rightarrow h^{1/\gamma} = e^{-1}$.

Thus, $\mu_{m,1} = \dfrac{1}{\gamma}$

$$\Rightarrow f_1(\mu_{m,1}) = \frac{h^{\mu_{m,1}}}{1!} \mu_{m,1}$$

$$= h^{1/\gamma}\left(\frac{1}{\gamma}\right) = \frac{e^{-1}}{\gamma} = \eta, \text{ say.} \tag{25}$$

In order to provide an estimate of $f_n(\mu_{m,n})$, we apply the A.M.-G.M. inequality on the n distinct positive quantities $\mu_{m,n}, (1+\mu_{m,n}), (2+\mu_{m,n}), \ldots, \{(n-1)+\mu_{m,n}\}$ to obtain:

$$\frac{1}{n}\left[\mu_{m,n} + (1+\mu_{m,n}) + (2+\mu_{m,n}) + \ldots + \{(n-1)+\mu_{m,n}\}\right]$$

$$> \left[\mu_{m,n}(1+\mu_{m,n})(2+\mu_{m,n})\ldots\{(n-1)+\mu_{m,n}\}\right]^{1/n}$$

$$\Rightarrow \frac{1}{n}\left[n\mu_{m,n} + \frac{n(n-1)}{2}\right] > \left[g_n(\mu_{m,n})\right]^{1/n}$$

$$\Rightarrow g_n(\mu_{m,n}) < \left[\mu_{m,n} + \frac{(n-1)}{2}\right]^n$$

$$\therefore f_{n,\max}(\mu)\Big|_{\beta > 0} = f_n(\mu_{m,n})$$

$$= \frac{h^{\mu_{m,n}}}{n!} g_n(\mu_{m,n}) < \frac{h^{\mu_{m,n}}}{n!}\left[\mu_{m,n} + \frac{(n-1)}{2}\right]^n$$

Putting back $\mu = -\alpha$ and $f_n(\mu) = \Phi_n(\alpha)$, we can now state that,

for $-\infty < \alpha < 0$, $\Phi_n(\alpha)$ rises monotonically from zero at $-\infty$ to a maximum at $\alpha = \alpha_{m,n} = -\beta_{m,n}$ and, with further increase in $\alpha$, it falls monotonically to zero at $\alpha = 0$, i.e. $\Phi_n(\alpha)$ is positive and bounded within 0 and some $\Phi_n(\alpha_{m,n})$.

The plots in Figs. 3 to 6 clearly illustrate the nature of $\Phi_n(\alpha)$ for negative $\alpha$.

$$\Phi_{n+1}(\alpha) = \frac{h^\beta}{(n+1)!}(-\alpha)(1-\alpha)(2-\alpha)\ldots\{(n-1)-\alpha\}(n-\alpha)$$

$$= \Phi_n(\alpha)\frac{(n-\alpha)}{n+1}$$

$$\therefore \Phi_{n+1}(\alpha) - \Phi_n(\alpha) = \Phi_n(\alpha)\left[\frac{(n-\alpha)}{n+1} - 1\right] = \Phi_n(\alpha)\left[\frac{-\alpha - 1}{n+1}\right].$$

As, for $\alpha < 0$, $\Phi_n(\alpha) > 0$, for any positive integer n, we conclude that $\Phi_{n+1}(\alpha) < \Phi_n(\alpha)$ for $-1 < \alpha < 0$ but $\Phi_{n+1}(\alpha) > \Phi_n(\alpha)$ for $\alpha < -1 \; \forall n$, i.e. $\Phi_n(\alpha)$ decreases with increase in n for any $-1 < \alpha < 0$ but increases with increase in n for any $\alpha < -1$. (26)

It is interesting to note that at $\alpha = -1$, $\Phi_n(-1) = \dfrac{h^1}{n!}(1)(1+1)(2+1)\ldots(n-1+1)$

$$= \frac{h}{n!}(n!) = h \; \forall n = 1,2,3,\ldots\ldots$$

so that at $\alpha = -1$, a crossover in the nature of $\Phi_n(\alpha)$ may be said to occur.

This means that, in the interval $-1 < \alpha < 0$, the maximum possible value of $\Phi_n(\alpha)$ occurs for $n = 1$ and is given by $\Phi_1(\alpha_{m,1}) = \eta$.

Again, $\dfrac{\Phi_{n+1}(\alpha) - \Phi_n(\alpha)}{\Phi_n(\alpha)} = \dfrac{-\alpha - 1}{n + 1} \to 0$ as $n \to \infty$, for any finite $\alpha$.

This implies that the relative change in the value of $\Phi_n(\alpha)$, for a given $\alpha$, becomes smaller and smaller as n becomes larger and larger. (27)

Now, getting back to $D^\alpha x(t) = \sum_{j=0}^{N} \Phi_j(\alpha) x(t - jh)$, we reiterate that, for any negative $\alpha$, the weights $\Phi_j(\alpha)$ are all positive.

For $-1 < \alpha < 0$, the value of $\Phi_j(\alpha)$ decreases as j increases, vide (26), and this value for any j can never exceed $\eta$, vide (25). In our case, we take the sampling interval as 0.001 secs so that $\gamma = -\ln(h) = -\ln(0.001) \approx 6.908$.

Thus, η = exp(–1)/γ ≈ 0.0532 << 1.

For any j, e(t – jh) is itself very small in magnitude so that multiplication with $\Phi_j(\alpha)$ will make it even smaller without altering its sign. Finally, as the large number (10000 in our case) of error-values is randomly distributed in the small interval $[-e_{max}, e_{max}]$ where $e_{max} << 1$, positive and negative values occur with nearly the same frequency. This discussion leads to the obvious conclusion that the net effect of the superposition of all these weighted error-values, which is $D^\alpha e(t)$, can be expected to be very small for $-1 < \alpha < 0$.

For $\alpha < -1$, it is true that the value of $\Phi_j(\alpha)$ increases as j increases but actual plots show that, for j = 0, 1, 2, …, 10000, this value is never too large; e.g., for negative α, the maximum value of $\Phi_{1000}(\alpha)$ is 2.80 which occurs at $\alpha \approx -10.5$. In fact, for smaller values of j, the weights $\Phi_j(\alpha)$ are very small (<< 1); for higher values of j, the weights are relatively larger but, as the variation in the value of $\Phi_j(\alpha)$ decreases with increasing j, vide (27), the values of $\Phi_j(\alpha)$ within a reasonable range of j are more or less comparable; e.g. at α = –4, $\Phi_j(\alpha)$ changes gradually from 0.0360 to 0.0379 (only about 5% change) as j changes from 6000 to 6100. However, within this interval, the error waveform has 101 samples (random) so that there are more or less evenly distributed positive and negative error-values and, even after multiplication with the respective weights, they will tend to cancel each other, on the average. So, once again, the overall weighted sum is likely to be small.

However, for our system, the output c(t) is always positive and may assume values greater than unity. Moreover, as all the weights are positive, the weighted sample values (however small owing to the smallness of the weights) will go on accumulating so that the weighted sum may be expected to be substantial.

Thus, we may be justified in expecting that, usually, $|D^\alpha e(t)| << |D^\alpha c(t)|$ for any negative α.

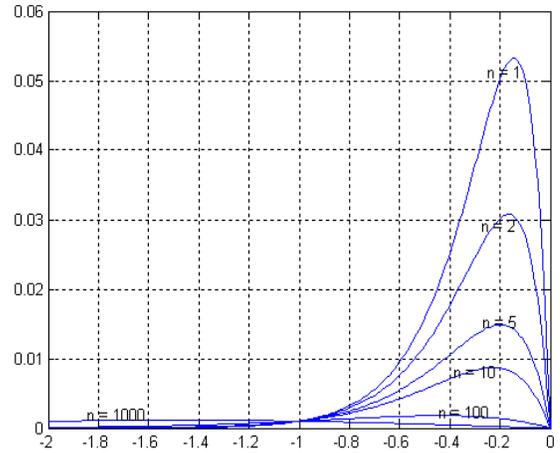

Figure 4. Plots of $\Phi_n(\alpha)$ vs. α for α < 0, n = 1, 2, 5, 10, 100, 1000.

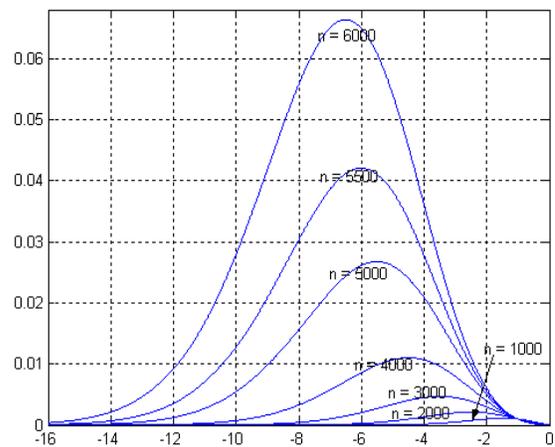

Figure 5. Plots of $\Phi_n(\alpha)$ vs. α for α < 0, n = 1000, 2000, 5000, 5500, 6000.

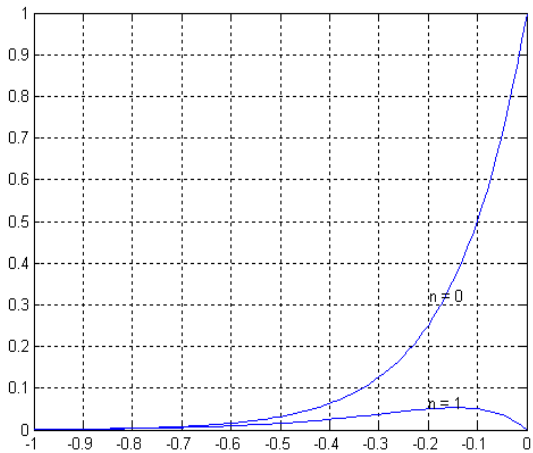

Figure 3. Plots of $\Phi_n(\alpha)$ vs. α for α < 0, n = 0, 1.

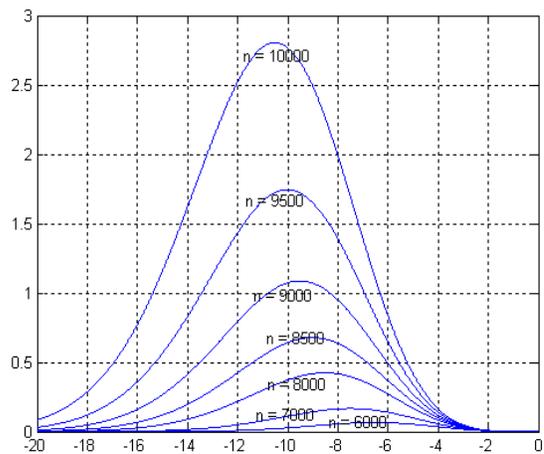

Figure 6. Plots of $\Phi_n(\alpha)$ vs. α for α < 0, n = 6000, 7000, 8000, 8500, 9000, 9500, 10000.

## B. α ≥ 0 : Fractional Order Differentiation

Let us write $\Phi_n(\alpha) = \dfrac{h^{-\alpha}}{n!} q(\alpha)$ where $q(\alpha) = (-\alpha)(1-\alpha)(2-\alpha)\ldots\{(n-1)-\alpha\}$ for some given n. Now, for $-\infty < \alpha < \infty$, $h^{-\alpha}$ ($0 < h < 1$) is always positive and is a strictly monotonically increasing function of α and, for α ≥ 0, the rate of increase is high (exponential). But, q(α) has n zeros at 0, 1, 2, …, n–1 so that, over the interval $0 \le \alpha \le n-1$, q(α) oscillates on both sides of the α–axis; in fact, as $q(\alpha) > 0 \,\forall \alpha < 0$ and $\alpha = 0$ is a zero-crossing point of q(α), we must have q(α) < 0 for $2m < \alpha < 2m+1$ and q(α) > 0 for $2m+1 < \alpha < 2m+2$, m = 0, 1, 2, ….., $\dfrac{n-2}{2}$ or $\dfrac{n-3}{2}$ according as (n –1) is odd or even.

This discussion suggests that, within $0 \le \alpha \le n-1$, the function $\Phi_n(\alpha) = \dfrac{h^{-\alpha}}{n!} q(\alpha)$ is oscillatory in nature but is modulated by a growing exponential envelope $\dfrac{h^{-\alpha}}{n!}$; for $(n-1) < \alpha < \infty$, $\Phi_n(\alpha)$ monotonically and sharply increases to $+\infty$ or decreases to $-\infty$ according as (n –1) is odd or even; i.e. it is unbounded beyond its last zero. However, there do exist some positive values of α at which the functional values are less than those at certain negative values of α.

Actual plots (Figs. 7 to 10) show that at certain non-negative values of α, the magnitude of $\Phi_j(\alpha)$ may be extremely large for smaller values of j although, roughly from j = 10 onwards, the value becomes much smaller for any α. So, it is not unlikely that, in the weighted sum, the terms corresponding to lower values of j will outweigh those corresponding to higher values of j. Moreover, for positive α, the weights $\Phi_j(\alpha)$ no longer have the same sign for all j so that signs of the terms in the weighted summation are no longer determined solely by e(t) or c(t). So, $D^\alpha e(t)$ may work out to a considerable value and may even be comparable to $D^\alpha c(t)$. In short, for non-negative α, we cannot ascribe any uniform pattern to the weights $\Phi_j(\alpha)$ and thus cannot make any generalized and conclusive claims about the relative magnitudes of $D^\alpha e(t)$ and $D^\alpha c(t)$ for any given t.

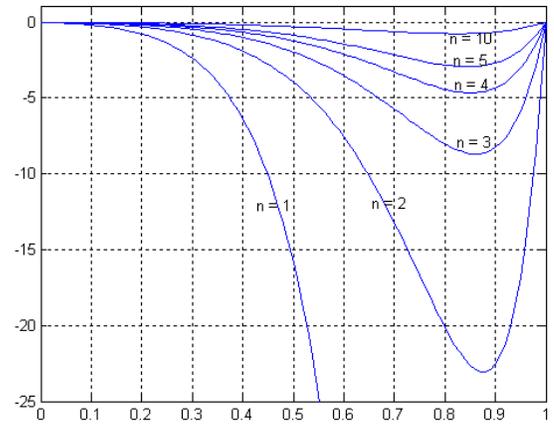

Figure 8. Plots of $\Phi_n(\alpha)$ vs. α for $0 < \alpha < 1$, n = 1, 2, 3, 4, 5, 10.

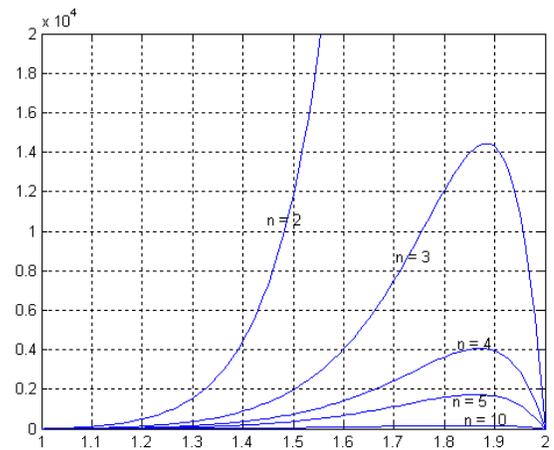

Figure 9. Plots of $\Phi_n(\alpha)$ vs. α for $1 < \alpha < 2$, n = 2, 3, 4, 5, 10.

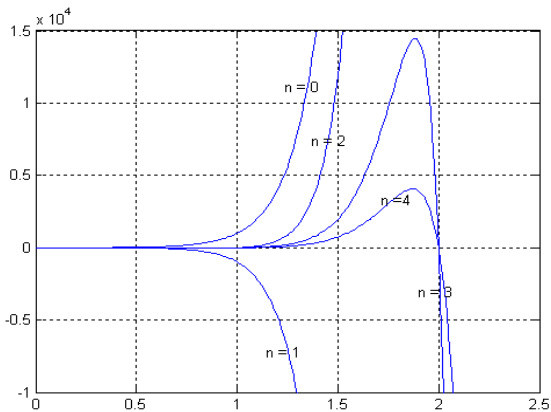

Figure 7. Plots of $\Phi_n(\alpha)$ vs. α for α > 0, n = 0, 1, 2, 3, 4.

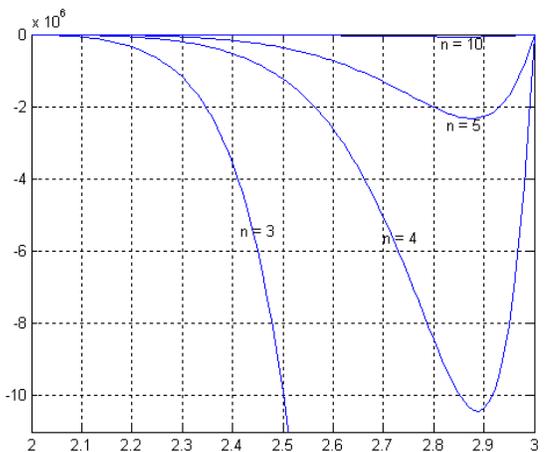

Figure 10. Plots of $\Phi_n(\alpha)$ vs. α for $2 < \alpha < 3$, n = 3, 4, 5, 10.

TABLE I. VARIATION OF $D^\alpha e(t)$ WITH $\alpha$. (THE 5 SEQUENCES $e(t)$ ARE CONSECUTIVE AND INDEPENDENT.)

| e(t) | $D^\alpha e(t)$ for derivation order $\alpha$ | | | | | | | | | |
|---|---|---|---|---|---|---|---|---|---|---|
| | $\alpha = 1.5$ | $\alpha = 1.2$ | $\alpha = 0.9$ | $\alpha = 0.6$ | $\alpha = 0.3$ | $\alpha = -0.3$ | $\alpha = -0.6$ | $\alpha = -0.9$ | $\alpha = -1.2$ | $\alpha = -1.5$ |
| 1 | -435.7842 | -50.7575 | -5.7583 | -0.6287 | -0.0661 | -0.0008 | 0.0001 | 0.0006 | 0.0013 | 0.0025 |
| 2 | -603.6659 | -59.5517 | -5.7933 | -0.5742 | -0.0617 | -0.0013 | -0.0005 | -0.0004 | -0.0002 | 0.0002 |
| 3 | 424.4136 | 44.8209 | 4.7948 | 0.5242 | 0.0581 | 0.0002 | -0.0002 | -0.0001 | -0.0001 | -0.0003 |
| 4 | -256.3730 | -26.5634 | -3.0495 | -0.3928 | -0.0549 | -0.0011 | -0.0002 | -0.0001 | -0.0002 | -0.0005 |
| 5 | -107.8138 | -12.0636 | -1.4119 | -0.1631 | -0.0164 | 0.0004 | 0.0004 | 0.0005 | 0.0008 | 0.0013 |

## VII. CONCLUSION

As an experimental validation of our thesis, we present in Table I, the values of $D^\alpha e(t)$ for 5 different sets of randomly generated e(t) with $\alpha = 1.5, 0.9, 0.3, –0.3, –0.9, –1.5$. The amplitude of e(t) varies between –0.01 and 0.01. Length of memory L = 10 seconds. Sampling rate is once in 0.001 seconds.

In the light of section VI, we contend that, to minimize the effect of error, we must ensure that all the fractional differintegrals to be computed are fractional integrations rather than fractional differentiations. To ensure this condition, we will need to apply suitable transformations on the original FODEs to obtain equivalent FODEs. This is illustrated with the help of the example of section IV where the system transfer function is of the form $\frac{1}{a_1 s^\alpha + a_2 s^\beta + a_3}$, and $\alpha, \beta > 0$. We can perform a simple transformation on the transfer function of the system, which we can write as $\frac{s^{-n}}{a_1 s^{\alpha-n} + a_2 s^{\beta-n} + a_3 s^{-n}}$, where $(n-1) < \alpha < n$ and $\alpha > \beta$. In this specific case, of course, $\alpha = 2.23, \beta = 0.88$ and $n = 3$. Using the modified set of simultaneous equations, in presence of the exact random error waveform (with its elements varying between –0.05 and 0.05) that was used for the experiment mentioned in section IV, we have $a_1 = 0.7992, a_2 = 0.4996, a_3 = 0.9996$ as the estimates of the unknown parameters. The errors in estimating them are respectively 0.1000%, 0.0800% and 0.0400%. Comparing this result with that obtained under ideal conditions, we can confidently say that the proposed scheme successfully eliminates the effects of noise.